# ON CALIBRATION OF DESIGN WEIGHTS

**SARJINDER SINGH[1*] AND RAGHUNATH ARNAB[2]**

**Summary**

In the present investigation, we build a bridge between the generalized regression (GREG) estimator due to Deville and Sarndal (1992) and the linear regression estimator due to Hansen, Hurwitz and Madow (1953) in the presence of single auxiliary variable. The bridge confirms that the sum of calibrated weights should be equal to sum of design weights as pointed out by Singh (2003, 2004, 2006) and Stearns and Singh (2008). An important modification in the statistical packages such as GES, SUDAAN etc. has been suggested.

**Keywords:** Calibration; Linear regression estimator; GREG; Estimation of total and variance.

-------------------
*-Author to whom correspondence should be addressed.
1. Department of Statistics and Computer Networking, St. Cloud State University, St. Cloud, MN 56301-4498, USA
2. Department of Statistics, University of Botswana, Botswana.

**Acknowledgements**: The authors' are also thankful to a professional English Editor Ms. Melissa Lindsey, Write Place Center, St. Cloud State University, for help in editing the entire manuscript.

## 1. Introduction

Deville and Särndal (1992) proposed the method of calibration estimators using auxiliary information. The proposed calibrated estimators provide a class of estimators. Some of the well-known estimators such as classical ratio-estimator belong to this class. Several authors including Singh (2003, 2004, 2006), Farrell and Singh (2002, 2005), Wu and Sitter (2001), Estevao and Särndal (2003), Kott (2003) and Montanari and Ronalli (2005) among others considered the Deville and Särndal (1992) method and derived important calibrated estimators. But, so far derivation of the traditional linear regression estimators from the class of calibrated estimators derived by Deville and Särndal (1992) method has not been found in the literature. In this present paper we have considered a subclass of the class of calibrated estimators provided by Deville and Särndal (1992). In this proposed subclass, the sum of calibrated weights remains equal to the sum of design weights as pointed out by Singh (2003, 2004, 2006) and Stearns and Singh (2008). The traditional regression estimator is found to belong to the proposed sub class.

Consider a finite population $\Omega = \{1, 2, ..., i, ..., N\}$ of $N$ units, from which a probability sample $s(s \subset \Omega)$ of fixed size $n$ is drawn with probability $p(s)$ according to a given sampling design $p$. The inclusion probabilities $\pi_i = \Pr(i \in s)$ and $\pi_{ij} \in \Pr(i \neq j \in s)$ are assumed to be strictly positive and known. Let $y_i$ be the value of the variable of interest, $y$, for the $i^{th}$ unit of the population, with which is also associated an auxiliary variable $x_i$. For the element $i \in s$, we observe $(y_i, x_i)$. The population total of the auxiliary variable $x$, $X = \sum_{i \in \Omega} x_i$, is assumed to be known. The objective is to estimate the population total $Y = \sum_{i \in \Omega} y_i$. Deville and Särndal (1992) proposed the calibrated estimator:

$$\hat{Y}_{ds} = \sum_{i \in s} w_i y_i \qquad (1.1)$$

for the Horvitz and Thompson (1952) estimator:

$$\hat{Y}_{HT} = \sum_{i \in s} \frac{y_i}{\pi_i} = \sum_{i \in s} d_i y_i \qquad (1.2)$$

where $d_i = 1/\pi_i$ and the calibrated weights $w_i$, $i \in s$ are obtained by minimizing chi-square type distance function:

$$\sum_{i \in s} \frac{(w_i - d_i)^2}{d_i q_i} \tag{1.3}$$

subject to the calibration constraint:

$$\sum_{i \in s} w_i x_i = X \tag{1.4}$$

Here $q_i$, $i \in s$ are suitably chosen weights. In many situations the value of $q_i = 1$. The form of the estimator (1.1) depends upon the choice of $q_i$. Minimization of (1.3) subject to calibration equation (1.4), leads to the calibrated weight:

$$w_i = d_i + \frac{d_i q_i x_i}{\sum_{i \in s} d_i q_i x_i^2} \left( X - \sum_{i \in s} d_i x_i \right) \tag{1.5}$$

Substitution of the value of $w_i$ from (1.5) in (1.1) leads to the generalized regression (GREG) estimator of the population total Y as:

$$\hat{Y}_{\text{GREG}} = \sum_{i \in s} d_i y_i + \hat{\beta}_{ds} \left( X - \sum_{i \in s} d_i x_i \right) \tag{1.6}$$

where

$$\hat{\beta}_{ds} = \left( \sum_{i \in s} d_i q_i x_i y_i \right) \bigg/ \left( \sum_{i \in s} d_i q_i x_i^2 \right) \tag{1.7}$$

An approximate variance of the calibrated estimator $\hat{Y}_{\text{GREG}}$ for a large sample size provided by Deville and Särndal (1992) as:

$$V_{\text{DS}}\left(\hat{Y}_{\text{GREG}}\right) = \frac{1}{2} \sum_{i \neq} \sum_{j \in \Omega} D_{ij} \pi_{ij} \left( d_i E_i - d_j E_j \right)^2 \tag{1.8}$$

where $D_{ij} = (\pi_i \pi_j - \pi_{ij})/\pi_{ij}$, $E_i = y_i - B x_i$ and $B = \sum_{i \in \Omega} q_i x_i y_i / \sum_{i \in \Omega} q_i x_i^2$.

A consistent and approximate unbiased estimator of variance proposed by them is:

$$\hat{V}_{\text{DS}}\left(\hat{Y}_{\text{GREG}}\right) = \frac{1}{2} \sum_{i \in s} \sum_{j \in s} D_{ij} \left( w_i e_i - w_j e_j \right)^2 \tag{1.9}$$

with $e_i = y_i - \hat{\beta}_{ds} x_i$.

## 2. Hansen, Hurwitz and Madow (1953) estimator using calibration

Deville and Särndal (1992) imposed the constraint $\sum_{i \in s} w_i x_i = X$ under the assumption that the value of the calibrated estimator $\hat{Y}_{ds} = \sum_{i \in s} w_i y_i$ for the total Y should be equal to the known total X if $y_i$ is replaced by $x_i$. In this section we find how one can derive the ordinary linear regression estimator using the calibrating weights derived by Deville and Särndal (1992).

Let us substitute:

$$q_i = q_i^* \left( \frac{\sum_{i \in s} d_i q_i^*}{\sum_{i \in s} d_i q_i^* x_i} - \frac{1}{x_i} \right) \quad (2.1)$$

in the expression (1.5). The substitution yields:

$$w_i = w_{i0} = d_i + \frac{d_i q_i^* \left( x_i \sum_{i \in s} d_i q_i^* - \sum_{i \in s} d_i q_i^* x_i \right)}{\left( \sum_{i \in s} d_i q_i^* x_i^2 \right)\left( \sum_{i \in s} d_i q_i^* \right) - \left( \sum_{i \in s} d_i q_i^* x_i \right)^2} \left( X - \sum_{i \in s} d_i x_i \right) \quad (2.2)$$

Finally putting $w_i = w_{i0}$ in the equation (1.1), we get:

$$\hat{Y}_{ds} = \hat{Y}_{LR} = \hat{Y}_{HT} + \hat{\beta}_{ols}\left(X - \hat{X}_{HT}\right) \quad (2.3)$$

with

$$\hat{\beta}_{ols} = \frac{\left(\sum_{i \in s} d_i q_i^*\right)\left(\sum_{i \in s} d_i q_i^* x_i y_i\right) - \left(\sum_{i \in s} d_i q_i^* y_i\right)\left(\sum_{i \in s} d_i q_i^* x_i\right)}{\left(\sum_{i \in s} d_i q_i^*\right)\left(\sum_{i \in s} d_i q_i^* x_i^2\right) - \left(\sum_{i \in s} d_i q_i^* x_i\right)^2}$$

It should be worth noting that the calibrated weights $w_{i0}$, $i \in s$ satisfy the constraints:

$$\sum_{i \in s} w_i x_i = X \quad (2.4)$$

and

$$\sum_{i \in s} w_i = \sum_{i \in s} d_i \quad (2.5)$$

Note that the condition (2.5) is due to Singh (2003, 2004, 2006). It builds a bridge between the GREG due to Deville and Sarndal (1992) and the linear regression estimator due to Hansen, Hurwitz and Madow (1953). Asymptotic properties of the estimator (2.3) are studied by Sampath and Chandra (1990). It reconfirms that there is a strong need to set constraint (2.5) into all the statistical packages like GES, SUDDAN etc. while doing calibration of design weights. Note that for simple random and without replacement (SRSWOR) sampling, the Wu and Sitter (2001), and Estevao and Sarndal (2003) calibration constraint is also a special case of (2.5) for $d_i = N/n$. For SRSWOR sampling $\pi_i = n/N$, the estimator (2.3) reduces to:

$$\hat{Y}_{LR} = N\left[\bar{y}_s + \hat{\beta}_{ols}\left(\bar{X} - \bar{x}_s\right)\right] \quad (2.6)$$

where $\bar{y}_s = \sum_{i \in s} y_i/n$, $\bar{x}_s = \sum_{i \in s} x_i/n$, and $\bar{X} = \sum_{i \in s} x_i/N$.

Further in particular $q_i^* = 1$, $\hat{\beta}_{ols}$ reduces to $s_{xy}/s_x^2$ and we get:

$$\hat{Y}_{LR} = N\left[\bar{y} + \frac{s_{xy}}{s_x^2}\left(\bar{X} - \bar{x}\right)\right] \quad (2.7)$$

where $s_x^2 = (n-1)^{-1}\sum_{i=1}^{n}(x_i - \bar{x})^2$ and $s_{xy} = (n-1)^{-1}\sum_{i=1}^{n}(x_i - \bar{x})(y_i - \bar{y})$. The estimator (2.6) is the famous traditional linear regression estimator due to Hansen, Hurwitz and Madow (1953) in the presence of a single auxiliary variable.

It is more interesting to note that Deville and Särndal (1992) assumed the model $M_1 : y_i = \beta x_i + e_i$ such that $e_i \overset{iid}{\sim} N(0, \sigma^2 v(x_i))$, with $v(x_i)$ being any function of $x_i$. Under model $M_1$, an estimator of $\beta$ can be obtained by $\min_{i \in s} \Sigma \, d_i q_i \hat{e}_i^2$ where $\hat{e}_i = y_i - \hat{\beta}_{ds}x_i$. In contrast the proposed method relaxes the condition of zero intercept, which is a requirement of the traditional linear regression estimator to provide efficient results, that is we can consider any linear model of the form $M_2 : y_i = \alpha + \beta x_i + e_i^*$, where $\alpha$ is intercept and $\beta$ is a slope, such that $e_i^* \overset{iid}{\sim} N(0, \sigma^2 v(x_i))$. Under model $M_2$, again the estimates of intercept and slope are given by $\min_{i \in s} \Sigma \, d_i q_i^* \hat{e}_i^{*2}$, where $\hat{e}_i^* = y_i - \hat{\alpha}_{ols} - \hat{\beta}_{ols}x_i$.

Simply put the new method relaxes the assumption of Deville and Särndal (1992) that the linear regression should pass through the origin, and the investigator or researcher need not be concerned about the status of the regression line while applying the proposed methodology.

## 3. Estimation of variance

Following Singh, Horn and Yu (1998), a calibrated estimator of the variance of the linear regression estimator $\hat{Y}_{LR}$ in (2.3) is given by:

$$\hat{V}_s(\hat{Y}_{LR}) = \frac{1}{2} \sum_{i \neq j \in s} \sum D_{ij} \Phi_{ij}(s) \qquad (3.1)$$

where $\Phi_{ij}(s) = (w_{i0}\hat{e}_i^* - w_{j0}\hat{e}_j^*)^2$ and $\hat{e}_i^*$ can be obtained by: $\min. \sum_{i \in s} d_i q_i^* \hat{e}_i^{*2}$. Further, we consider a new calibrated estimator of variance of the linear regression as:

$$\hat{V}_{ss}(\hat{Y}_{LR}) = \frac{1}{2} \sum_{i \neq j \in s} \sum \Omega_{ij}(s) \Phi_{ij}(s) \qquad (3.2)$$

where $\Omega_{ij}(s)$ are weights such that the chi-square distance function:

$$D = \frac{1}{2} \sum_{i \neq j \in s} \sum \frac{\left(\Omega_{ij}(s) - D_{ij}\right)^2}{D_{ij} Q_{ij}(s)} \qquad (3.3)$$

is minimum subject to a calibration constraint, given by:

$$\frac{1}{2} \sum_{i \neq j \in s} \sum \Omega_{ij}(s) \delta_{ij} = V_{syg}(\hat{X}_{HT}) \qquad (3.4)$$

where $V_{syg} = \frac{1}{2} \sum_{i \neq j \in \Omega} \sum D_{ij} \pi_{ij} \delta_{ij}$ and $\delta_{ij} = (d_i x_i - d_j x_j)^2$. Obviously, for the minimization of (3.3) subject to (3.4), the Lagrange function is given by:

$$LM = \frac{1}{2} \sum_{i \neq j \in s} \sum \frac{\left(\Omega_{ij}(s) - D_{ij}\right)^2}{D_{ij} Q_{ij}(s)} - \mu \left\{ \frac{1}{2} \sum_{i \neq j \in s} \sum \Omega_{ij}(s) \delta_{ij} - V_{syg}(\hat{X}_{HT}) \right\} \qquad (3.5)$$

with $\mu$ as a Lagrange multiplier. On setting $\partial LM / \partial \Omega_{ij}(s) = 0$, we have:

$$\Omega_{ij}(s) = D_{ij} + \mu \frac{D_{ij} Q_{ij}(s)}{2} \delta_{ij} \qquad (3.6)$$

On using (3.4) in (3.5) we have:

$$\mu = 4 \left\{ V_{syg}(\hat{X}_{HT}) - \hat{V}_{syg}(\hat{X}_{HT}) \right\} \Big/ \sum_{i \neq j \in s} \sum D_{ij} Q_{ij}(s) \delta_{ij}^2$$

and

$$\Omega_{ij}(s) = D_{ij} + \frac{2 D_{ij} Q_{ij}(s) \delta_{ij} \left\{ V_{syg}(\hat{X}_{HT}) - \hat{V}_{syg}(\hat{X}_{HT}) \right\}}{\sum_{i \neq j \in s} \sum D_{ij} Q_{ij}(s) \delta_{ij}^2} \qquad (3.7)$$

noting $\hat{V}_{syg} = \frac{1}{2}\sum_{i \neq j \in s}\sum D_{ij}\delta_{ij}$ denotes the Sen (1953) and Yates and Grundy (1953) form of the estimator of variance. On substituting (3.7) into (3.2), we obtain a new calibrated estimator of variance of the linear regression estimator $\hat{Y}_{LR}$ in (2.3) as:

$$\hat{V}_{ss}(\hat{Y}_{LR}) = \hat{V}_{s}(\hat{Y}_{LR}) + \hat{B}_2\{V_{syg}(\hat{X}_{HT}) - \hat{V}_{syg}(\hat{X}_{HT})\} \tag{3.8}$$

where

$$\hat{B}_2 = \frac{\sum_{i \neq j \in s}\sum D_{ij}Q_{ij}(s)\delta_{ij}\Phi_{ij}(s)}{\sum_{i \neq j \in s}\sum D_{ij}Q_{ij}(s)\delta_{ij}^2} \tag{3.9}$$

Now choosing:

$$Q_{ij}(s) = Q_{ij}^{*}(s) = q_{ij}(s)\left[\frac{\sum_{i \neq j \in s}\sum D_{ij}q_{ij}(s)}{\sum_{i \neq j \in s}\sum D_{ij}q_{ij}(s)\delta_{ij}} - \frac{1}{\delta_{ij}}\right] \tag{3.10}$$

with $q_{ij}(s)$ as a suitable weight we get:

$$\hat{B}_2 = \frac{\left(\sum_{i \neq j \in s}\sum D_{ij}Q_{ij}(s)\delta_{ij}\Phi_{ij}(s)\right)\left(\sum_{i \neq j \in s}\sum D_{ij}q_{ij}(s)\right) - \left(\sum_{i \neq j \in s}\sum D_{ij}q_{ij}(s)\delta_{ij}\right)\left(\sum_{i \neq j \in s}\sum D_{ij}q_{ij}(s)\Phi_{ij}(s)\delta_{ij}\right)}{\left(\sum_{i \neq j \in s}\sum D_{ij}q_{ij}(s)\delta_{ij}^2\right)\left(\sum_{i \neq j \in s}\sum D_{ij}q_{ij}(s)\right) - \left(\sum_{i \neq j \in s}\sum D_{ij}q_{ij}(s)\delta_{ij}\right)^2} \tag{3.11}$$

The choice of $Q_{ij}^{*}(s)$ in (3.10) satisfies constraints:

$$\frac{1}{2}\sum_{i \neq j \in s}\sum \Omega_{ij}^{*}(s)\delta_{ij} = V_{syg}(\hat{X}_{HT}) \tag{3.12}$$

and

$$\sum_{i \neq j \in s}\sum \Omega_{ij}^{*}(s) = \sum_{i \neq j \in s}\sum D_{ij} \tag{3.13}$$

Again note that the condition (3.13) is due to Singh (2003, 2004, 2006). Thus it builds a bridge between the estimator of variance due to Singh, Horn and Yu (1998) and Singh (2003, 2004, 2006).

**Remark:** Note carefully if $w_i^* = d_i$ and $e_i^* = y_i$, then the estimator $\hat{Y}_{LR} = \hat{Y}_{HT}$, then following Singh, Horn, Yu and Chowdhury (1999), the ratio $\hat{V}_{ss}(\hat{Y}_{LR})/\{N^2(1-f)/n\}$

becomes a traditional linear regression estimator of finite population variance, $\sigma_y^2 = N^{-1}\sum_{i=1}^{N}(Y_i - \bar{Y})^2$, under SRSWOR sampling given by:

$$\hat{\sigma}_{dt}^2 = s_y^2 + \hat{\beta}_2\left(S_x^2 - s_x^2\right) \tag{3.14}$$

where $s_y^2 = (n-1)^{-1}\sum_{i=1}^{n}(y_i - \bar{y})^2$, $S_x^2 = (N-1)^{-1}\sum_{i=1}^{N}(X_i - \bar{X})^2$, $\hat{\beta}_2 = \left(\hat{\mu}_{22} - \hat{\mu}_{20}\hat{\mu}_{02}\right)/\left(\hat{\mu}_{04} - \hat{\mu}_{02}^2\right)$

with $\hat{\mu}_{rs} = (n-1)^{-1}\sum_{i\in s}(y_i - \bar{y})^r (x_i - \bar{x})^s$, which was obtained by Das and Tripathi (1978). Note that the estimator (3.14) has also independently studied by Srivastava and Jhajj (1980) and Isaki (1983).

## 4. Stratified sampling design

Suppose that the population consists of $L$ strata with $N_h$ units in the $h^{th}$ stratum from which a simple random sample of size $n_h$ is taken without replacement, then the total population size $N = \sum_{h=1}^{L} N_h$ and sample size $n = \sum_{h=1}^{L} n_h$. Let the $i^{th}$ unit of the $h^{th}$ stratum be associated with two values $y_{h_i}$ and $x_{h_i}$ with $x_{h_i} > 0$ being the covariate. For the $h^{th}$ stratum, let $W_h = N_h/N$ be the stratum weights, $f_h = n_h/N_h$ the sample fraction, $\bar{y}_h$, $\bar{x}_h$, $\bar{Y}_h$, $\bar{X}_h$ the $y$ and $x$ sample and population means respectively. Assume $\bar{X} = \sum_{h=1}^{L} W_h \bar{X}_h$ is known. The purpose is to estimate $\bar{Y} = \sum_{h=1}^{L} W_h \bar{Y}_h$, possibly by incorporating the covariance information $x$. The usual estimator of population mean $\bar{Y}$ is given by:

$$\bar{y}_{st} = \sum_{h=1}^{L} W_h \bar{y}_h \tag{4.1}$$

Singh, Horn and Yu (1998) and Tracy, Singh and Arnab (2003) considered an estimator, given by:

$$\bar{y}_{St}^* = \sum_{h=1}^{L} W_h^* \bar{y}_h \tag{4.2}$$

with new weights $W_h^*$. The new weights $W_h^*$ are chosen such that chi square type distance given by:

$$\sum_{h=1}^{L} \frac{\left(W_h^* - W_h\right)^2}{W_h q_h} \tag{4.3}$$

is minimum subject to the condition:

$$\sum_{h=1}^{L} W_h^* \bar{x}_h = \bar{X} \tag{4.4}$$

Minimization of (4.3) subject to the calibration equation (4.4) leads to the combined regression type estimator given by:

$$\bar{y}_{St}^* = \sum_{h=1}^{L} W_h \bar{y}_h + \frac{\sum_{h=1}^{L} W_h q_h \bar{x}_h \bar{y}_h}{\sum_{h=1}^{L} W_h q_h \bar{x}_h^2} \left[ \bar{X} - \sum_{h=1}^{L} W_h \bar{x}_h \right] \tag{4.5}$$

Note that the estimator (4.5) suggested by Singh, Horn and Yu (1998) is not a traditional linear combined regression estimator.

## 5. Combined linear regression using calibration

We consider here a new estimator of the population mean $\bar{Y}$ in stratified sampling as:

$$\bar{y}_{St}^0 = \sum_{h=1}^{L} W_h^0 \bar{y}_h \tag{5.1}$$

where $W_h^0$ are the calibrated weights such that the chi square distance function:

$$D^0 = \frac{1}{2} \sum_{h=1}^{L} \frac{\left(W_h^0 - W_h\right)^2}{W_h Q_h^0} \tag{5.2}$$

is minimum subject to two constraints, defined as:

$$\sum_{h=1}^{L} W_h^0 = \sum_{h=1}^{L} W_h \tag{5.3}$$

and

$$\sum_{h=1}^{L} W_h^0 \bar{x}_h = \bar{X} \tag{5.4}$$

where $Q_h^0$ are some suitably chosen weights. The condition (5.3) implies that the sum of observed weights should be equal to the sum of expected weights across all strata. Thus, the new calibrated weights are given by:

$$W_h^0 = W_h + \frac{(W_h Q_h \bar{x}_h)\left(\sum_{h=1}^{L} W_h Q_h\right) - (W_h Q_h)\left(\sum_{h=1}^{L} W_h Q_h \bar{x}_h\right)}{\left(\sum_{h=1}^{L} W_h Q_h\right)\left(\sum_{h=1}^{L} W_h Q_h \bar{x}_h^2\right) - \left(\sum_{h=1}^{L} W_h Q_h \bar{x}_h\right)^2} \left( \bar{X} - \sum_{h=1}^{L} W_h \bar{x}_h \right) \tag{5.5}$$

and thus a new calibrated estimator of the population mean $\bar{Y}$ becomes:

$$\bar{y}_s^{st} = \sum_{h=1}^{L} W_h \bar{y}_h + \hat{\beta}_{st}\left[\bar{X} - \sum_{h=1}^{L} W_h \bar{x}_h\right] \tag{5.6}$$

where:

$$\hat{\beta}_{st} = \frac{\left(\sum_{h=1}^{L} W_h Q_h^0 \bar{x}_h \bar{y}_h\right)\left(\sum_{h=1}^{L} W_h Q_h^0\right) - \left(\sum_{h=1}^{L} W_h Q_h^0 \bar{y}_h\right)\left(\sum_{h=1}^{L} W_h Q_h^0 \bar{x}_h\right)}{\left(\sum_{h=1}^{L} W_h Q_h^0\right)\left(\sum_{h=1}^{L} W_h Q_h^0 \bar{x}_h^2\right) - \left(\sum_{h=1}^{L} W_h Q_h^0 \bar{x}_h\right)^2} \tag{5.7}$$

If $Q_h^0 = 1$ then the estimator (5.6) reduces to the traditional combined stratified linear regression estimator, and hence better than the estimators developed by Singh, Horn and Yu (1998), and Tracy, Singh and Arnab (2003).

## 6. Estimation of Variance of Combined Linear Regression

Now we consider a new estimator of the variance of the combined stratified linear regression estimator and correct the results of Singh, Horn and Yu (1998). The well-known estimator of variance of combined regression estimator is given by:

$$\hat{v}\left(\bar{y}_{St}^*\right) = \sum_{h=1}^{L} \frac{W_h^2(1-f_h)}{n_h} s_{e_h^*}^2 \tag{6.1}$$

where $s_{e_h^*}^2 = (n_h - 1)^{-1} \sum_{i=1}^{n_h} e_{hi}^{*2}$ is the $h^{th}$ stratum sample variance and $e_{hi}^* = (y_{hi} - \bar{y}_h) - b_{st}(x_{hi} - \bar{x}_h)$ and $b_{st}$ denote the traditional linear regression coefficient in stratified sampling.

The calibration approach yields an estimator of variance of the combined regression estimator as:

$$\hat{v}_c\left(\bar{y}_{St}^*\right) = \sum_{h=1}^{L} \frac{D_h(W_h^0)^2}{W_h^2} s_{e_h^*}^2 \tag{6.2}$$

where $D_h = \frac{W_h^2(1-f_h)}{n_h}$ and $W_h^0$ is given by (5.5). Again following Singh, Horn and Yu (1998), a calibration estimator of the variance of combined linear regression estimator is given by:

$$\hat{v}_{St}\left(\hat{Y}_s^{St}\right) = \sum_{h=1}^{L} \frac{\Omega_h^0 (W_h^0)^2}{W_h^2} s_{e_h^*}^2 \tag{6.3}$$

where $\Omega_h^0$ are suitably chosen weights such that chi square distance function given by:

$$\sum_{h=1}^{L} \frac{\left(\Omega_h^0 - D_h\right)^2}{D_h Q_h^0} \tag{6.4}$$

is minimum subject to two calibration equations defined as:

$$\sum_{h=1}^{L} \Omega_h^0 = \sum_{h=1}^{L} D_h \tag{6.5}$$

and

$$\sum_{h=1}^{L} \Omega_h^0 s_{hx}^2 = V(\bar{x}_{St}) \tag{6.6}$$

where $V(\bar{x}_{St}) = \sum_{h=1}^{L} W_h^2 \{(1-f_h)/n_h\} S_{hx}^2$ is assumed to be known, and $s_{hx}^2 = (n_h - 1)^{-1} \sum_{i=1}^{n_h} (x_{hi} - \bar{x}_h)^2$ is an unbiased estimator of $S_{hx}^2 = (N_h - 1)^{-1} \sum_{i=1}^{N_h} (X_{hi} - \bar{X}_h)^2$, and $\hat{v}(\bar{x}_{St}) = \sum_{h=1}^{L} W_h^2 \{(1-f_h)/n_h\} s_{hx}^2$ is an unbiased estimator of $V(\bar{x}_{St})$.

The calibrated weights are then given by:

$$\Omega_h^0 = D_h + \frac{\left(D_h Q_h^0 s_{hx}^2\right)\left(\sum_{h=1}^{L} D_h Q_h^0\right) - \left(D_h Q_h^0\right)\left(\sum_{h=1}^{L} D_h Q_h^0 s_{hx}^2\right)}{\left(\sum_{h=1}^{L} D_h Q_h^0\right)\left(\sum_{h=1}^{L} D_h Q_h^0 s_{hx}^4\right) - \left(\sum_{h=1}^{L} D_h Q_h^0 s_{hx}^2\right)^2} \{V(\bar{x}_{St}) - \hat{v}(\bar{x}_{St})\} \tag{6.7}$$

This procedure leads to a new calibrated estimator for the variance of the combined linear regression estimator given by:

$$\hat{v}\left(\hat{y}_s^{St}\right)_{ho} = \hat{v}\left(\hat{Y}_s^{St}\right) + \hat{B}_{St}^0 \left[V(\bar{x}_{St}) - \hat{v}(\bar{x}_{St})\right] \tag{6.8}$$

where

$$\hat{B}_{St}^0 = \frac{\left(\sum_{h=1}^{L} \frac{D_h Q_h^0 (W_h^0)^2 s_{hx}^2 s_{e*h}^2}{W_h^2}\right)\left(\sum_{h=1}^{L} D_h Q_h^0\right) - \left(\sum_{h=1}^{L} \frac{D_h Q_h^0 s_{hx}^2 s_{e*h}^2}{W_h^2}\right)\left(\sum_{h=1}^{L} D_h Q_h^0 s_{hx}^2\right)}{\left(\sum_{h=1}^{L} D_h Q_h^0\right)\left(\sum_{h=1}^{L} D_h Q_h^0 s_{hx}^4\right) - \left(\sum_{h=1}^{L} D_h Q_h^0 s_{hx}^2\right)^2} \tag{6.9}$$

which seems to be a completely new development although it is a corrected version of the estimator due to Singh, Horn and Yu (1998) and hence that of Tracy, Singh, and Arnab (2003).

# 7. Simulation Study

**7.1. Finite population:** In the case of finite populations, we have taken a population consisting of $N = 20$ units from Horvitz and Thompson (1952). The study variable, $y^*$, is the number of households on $i^{th}$ block and known auxiliary character, $x^*$ is the eye estimated number of households on the $i^{th}$ block. All possible samples of size $n = 5$ were selected by SRSWOR, which resulted in $\binom{N}{n} = 15{,}504$ samples. Then from the $k^{th}$ sample, the estimators $\hat{\bar{y}}_{LR}|_k = \bar{y} + \hat{\beta}_{ols}(\bar{X} - \bar{x})$ and $\hat{\bar{y}}_{ds}|_k = \bar{y} + \hat{\beta}_{ds}(\bar{X} - \bar{x})$ were computed. The empirical mean squared error of these estimators was computed as:

$$\text{MSE}(\hat{\bar{y}}_{LR}) = \binom{N}{n}^{-1} \sum_{k=1}^{\binom{N}{n}} [\hat{\bar{y}}_{lr}|_k - \bar{Y}]^2 \text{ and } \text{MSE}(\hat{\bar{y}}_{ds}) = \binom{N}{n}^{-1} \sum_{k=1}^{\binom{N}{n}} [\hat{\bar{y}}_{ds}|_k - \bar{Y}]^2$$

The percent relative efficiency of the proposed estimator with respect to the Deville and Särndal (1992) estimator was computed as:

$$\text{RE} = \frac{\text{MSE}(\hat{\bar{y}}_{ds})}{\text{MSE}(\hat{\bar{y}}_{LR})} \times 100. \tag{7.1}$$

The values of the relative efficiencies are presented in Table 1.

**Table 1.** Comparison of $\hat{\bar{y}}_{LR}$ with respect to $\hat{\bar{y}}_{ds}$

| Sr. No. | Transformations | | Correlation $\rho_{xy}$ | Sample Size | Relative Efficiency (%) |
|---|---|---|---|---|---|
| 1 | $y = \sqrt{y^*}$ | $x = x^*$ | 0.890 | 5 | 403.32 |
| | | | | 6 | 405.55 |
| | | | | 7 | 401.38 |
| 2 | $y = y^*$ | $x = \sqrt{x^*}$ | 0.867 | 5 | 122.23 |
| | | | | 6 | 129.23 |
| | | | | 7 | 131.77 |
| 3 | $y = \log(y^*)$ | $x = x^*$ | 0.897 | 5 | 1404.98 |
| | | | | 6 | 1423.30 |
| | | | | 7 | 1412.93 |
| 4 | $y = y^*$ | $x = \log(x^*)$ | 0.856 | 5 | 138.12 |
| | | | | 6 | 153.12 |
| | | | | 7 | 159.67 |

In order to study the different situations, we applied different transformations on the study variable $y^*$ and auxiliary variable $x^*$ which affect the value of the relationship between these two variables. The values of $\rho_{xy}$, sample sizes, transformations applied, and relative efficiency figures are given in Table 1. Note that the results given in Table 1 are exact and can be reproduced at any time, or a copy of the program can be had from the authors upon request. From the exact results given in Table 1 we can easily see that the gain in relative efficiency of the proposed technique is appreciable.

In real life situations, the study variable and auxiliary variables may follow specific kinds of distributions like normal, beta, or gamma etc. In order to see the performance of the proposed strategies under such circumstances, we generated artificial populations by following Singh, Horn and Yu (1998) and considered the problem of estimation of finite population mean through simulation as follows.

**7.2. Infinite populations:** The size $N$ of these populations is unknown. We generated a pair of $n$ independent random numbers $y_i^*$ and $x_i^*$ (say), $i = 1, 2, \ldots, n$, from a subroutine VNORM with PHI = 0.7, seed(y) = 13031963, and seed(x) = 19630313 following Bratley, Fox, and Scharge (1983). For fixed $S_y^2 = 50$ and $S_x^2 = 50$, we generated transformed variables:

$$y_i = 100.0 + \sqrt{S_y^2(1-\rho_{xy}^2)}\, y_i^* + \rho S_y x_i^* \qquad \text{and} \qquad x_i = 90.0 + S_x x_i^*$$

for different values of the correlation coefficient $\rho_{xy}$. For the $k^{th}$ sample, the estimators: $\hat{\bar{y}}_{LR}\vert_k = \bar{y} + \hat{\beta}_{ols}(\bar{X}-\bar{x})$ and $\hat{\bar{y}}_{ds}\vert_k = \bar{y} + \hat{\beta}_{ds}(\bar{X}-\bar{x})$ were computed. Then the empirical mean squared error of these estimators were, respectively, approximated as:

$$\text{MSE}(\hat{\bar{y}}_{LR}) = (15000)^{-1} \sum_{k=1}^{15000} \left[\hat{\bar{y}}_{LR}\vert_k - Y\right]^2 \quad \text{and} \quad \text{MSE}(\hat{\bar{y}}_{ds}) = (15000)^{-1} \sum_{k=1}^{15000} \left[\hat{\bar{y}}_{ds}\vert_k - Y\right]^2$$

The percent relative efficiency of the proposed estimator with respect to Deville and Särndal (1992) estimator was computed as:

$$\text{RE} = \frac{\text{MSE}(\hat{\bar{y}}_{ds})}{\text{MSE}(\hat{\bar{y}}_{LR})} \times 100 \tag{7.2}$$

The results so obtained are presented in Table 2.

**Table 2.** The percent RE of $\hat{\bar{y}}_{LR}$ with respect to $\hat{\bar{y}}_{ds}$

|             | Correlation coefficient ||||| 
| Sample Size | 0.1    | 0.3    | 0.5    | 0.7    | 0.9    |
|-------------|--------|--------|--------|--------|--------|
| 25          | 160.24 | 141.32 | 130.69 | 116.75 | 108.82 |
| 50          | 161.62 | 146.60 | 131.34 | 119.37 | 114.45 |
| 75          | 162.58 | 145.47 | 128.70 | 118.63 | 113.43 |
| 100         | 165.88 | 145.49 | 129.80 | 117.46 | 113.13 |

Table 2 illustrates that for moderate sample sizes the relative efficiency of the proposed technique also remains better than its competitors, and hence we have the following conclusion. If the value of correlation is high, say 0.9 then as the sample size becomes 50 the relative efficiency is maximum, and as the sample size becomes 100 the relative efficiency decreases, which makes sense because when the sample size approaches infinity then both estimators may be conversing to each other. For the low value of correlation coefficient, the relative efficiency of the proposed estimator is still continuing to increase as the same size increases, which again makes sense because the traditional linear regression estimator is always more efficient than the simple mean estimator irrespective of the non-zero magnitude of the correlation coefficient unlike the Deville and Särndal (1992) estimator. Hence more gains are expected for such situations with respect to Deville and Särndal (1992).